\documentclass[11pt]{article}
\usepackage{epsf}
\usepackage{epstopdf}
\begin{document}
\def\be{\begin{equation}}
\def\ee{\end{equation}}
\def\bee{\begin{eqnarray}}
\def\eee{\end{eqnarray}}
\def\ni{\noindent}
\def\bd{\begin{displaymath}}
\def\ed{\end{displaymath}}

\sloppy

\title {{\bf Degree distribution and scaling in the connecting - nearest - neighbors model}}

\vskip 1cm

\author{{ \bf Boris Rudolf} \\
Dept. of Mathematics\\ 
Faculty of El. Engineering  and Information
Technology, \\
Slovak Technical University, Bratislava, Slovakia
\and { \bf  M\'aria Marko\v sov\'a, Martin \v Caj\'agi,} \\
Dept. of Applied Informatics\\ 
Faculty of Mathematics, Physics and
Informatics \\
Comenius University, Bratislava, Slovakia
\and {\hskip 1truecm \bf Peter Ti\v no}\\
School of Computer Science\\ 
The University of Birmingham,
Birmingham, United Kingdom}
\maketitle

\begin{abstract}
We present a detailed analysis of the Connecting Nearest Neighbors (CNN)
model  by V\'azquez. We show that the degree distribution
follows a power law, but the scaling exponent can vary with the
parameter setting. Moreover, the correspondence of the growing version of the Connecting Nearest Neighbors (GCNN) model to the
particular random walk model (PRW model) and recursive search model (RS
model) is established.
\end{abstract}

\newpage
\section{Introduction}
It is well known, that the structure of growing networks is influenced by the dynamical processes of their creation such as node addition 
and edge linking \cite{Vas, My, BA, DM1, AB}.  
To show this explicitly, several models of  dynamically evolving networks have been proposed \cite{BA, DM1, DM3}. 

Many real networks (such as the Internet, for example) expand in time. The number of nodes added to 
the system  far exceeds the number of vanishing nodes.  Such networks are  well described by 
the growing  models \cite{DM1}.  The dynamics of the network growth is captured  by the  integro - differential 
dynamical equations. Some growing network models are analytically solvable, 
but often we have to rely on numerical simulations or  approximate solutions  \cite{Vas, BA, DM1, AB}. 
Numerical simulations of the growing network usually include the following processes:

\begin{itemize}
\item  at each time step a new node is added to the network, while  bringing $m$ new links

\item  nodes are labeled according to the time $s$ in which they joined the network

\item  each link attaches itself to the older nodes with certain probability, for example preferentially.
\end{itemize}

Barab\'asi and Albert \cite{BA} showed, both numerically and analytically, that the preferential node 
attachment leads to  scale free network structure. Preferential node attachment means that the linking probability 
of the edge from the newly added  node to an existing node is proportional to the receiving node degree. Ignoring edge direction, the node degree simply counts the number of 
edges incident with that node in the network.  The notion of ``scale free-ness'' signifies that there is no dominant scale 
defining the network, except of its size. Scale free networks have complex structure with degree  fluctuations 
of all sizes,  as manifested in the power law degree distribution

\be\label{degdist1}
P(k)\propto k^{-\gamma},
\ee
\ni where $P(k)$ denotes the probability of choosing\footnote{w.r.t. uniform probability over nodes in the network} a node with the degree $k$
 and  $\gamma>0$ is a scaling exponent depending on the network dynamics. It has been shown analytically, that 
for the Barab\'asi - Albert model $\gamma_{BA}=3.0$ \cite{BA}. If the node linking  differs from  the `pure' preferential 
one, $\gamma$ differs from $\gamma_{BA}$ \cite{Vas, DM1}, or even the network is no longer scale free.  For example, if the 
linking probability is random, degree distribution looses its power law character \cite{DM1, AB}.

There are several real examples of  scale free networks created by the self organizing processes. 
Some of them, such as the Internet at the autonomous system level, social network or language network \cite{VPV, CSSZ, Mark1} 
have even a scale free hierarchical structure.  Hierarchy in growing networks 
has been studied by Rav\'asz and Barab\'asi \cite{RB} and also by us  \cite{My}. It has been shown, both numerically and 
analytically, that the hierarchical 
node organization is reflected in the power law distribution of average clustering coefficients $C(k)$ of nodes 
with the degree $k$

\be\label{cldist1}
C(k)\propto k^{-\delta}.
\ee

\ni Here $\delta>0$ is a scaling exponent. Scale free hierarchical networks have power law degree and 
clustering coefficient distributions (\ref{degdist1}) and  (\ref{cldist1}), respectively. 

Pure preferential linking leads to scale free, but not to  hierarchical network structure. 
It has been shown recently  \cite{RB, My}, that both eqs. (\ref{degdist1}) and (\ref{cldist1}) hold, if the process of network 
growth involves  some local dynamics - the first  of the new $m$  links brought by a new node $s$ to the 
system is attached randomly with linking probability $1\over t$ (where $t\approx N$, the number of nodes), and the other 
$m-1$ links attach to the neighbors of the first choice node. The general scale free hierarchical model (SFH model) 
has been studied numerically   \cite{My, RB}.  For its simplified version with $m=2$  (SSFH model) we succeeded in finding scaling 
exponents $\gamma$ and $\delta$ analytically \cite{My}. 

A different framework for studying complex networks  has been presented by V\'azquez in \cite{Vas} - the network structure is  
discovered through random walk by several surfers. Such a random walk model  can be transformed to our SSFH model \cite{My}. 
Another  model by V\'azquez, namely the connecting nearest neighbors (CNN) model (section \ref{sec:CNN}), has been solved 
analytically using the idea of potential edges and potential degree of the node \cite{Vas}.  
Differential equations, describing the model dynamics, were treated using several approximations. 

In this paper we present a more accurate analysis of the the CNN model. We show, how the asymptotic degree distribution 
changes  when  the line of transcritical bifurcation in the parameter space is crossed. 
We relate  our more detailed analytical solution to the original solution by 
V\'azquez \cite{Vas}. We also show, that the CNN model corresponds  
(for certain parameter values)  to the other known models described in \cite{Vas}.

The paper has the following organization: In the second chapter we present the CNN model. Chapter $3$ is devoted to the qualitative analysis
of the nonlinear system of differential equations describing  the dynamics of the CNN network. In the fourth chapter 
the correspondence of the growing version of the CNN model (GCNN model \cite{Vas}) to the particullar random walk model (PRW model
 \cite{Vas}) and the recursive search model (RS model \cite{Vas})  is discussed. Section 5 concludes the paper by summarizing the main findings of this study.

\section{The CNN model}
\label{sec:CNN}
The CNN model  has been inspired by social networks \cite{Vas}. It is assumed that in such networks
 two sites with a common neighbor are connected with greater probability then two randomly chosen sites. 
This reflects the fact, that  in the social network it is more probable that two people (nodes) 
know themselves (are connected) if they have a common friend (common neighbor).
An analytical understanding of the CNN model has been achieved using the notions of potential edges and potential 
degree.

The dynamics of the CNN network is defined by the transition rates of link states from node $s$
(each possible link from $s$ to other nodes in the network can be in three possible states - disconnected ($d$),
potential edge ($p$), or an edge ($e$)).
For each node $s$ the transition rates $\nu_{x\rightarrow y}(s)$ 
$x,y \in \{d,p,e\}$ are defined per link.
A potential edge is defined as follows:  two nodes are connected
by a potential edge, if they are not connected by an edge and they
have at least one common neighbor. For each node $s$ in the growing
CNN network  a degree and a potential degree can  therefore be defined.

Consider a network with $N$ nodes. Denote the (real) degree and potential degree of a node $s$ by $k(s,N)$ and $k^*(s,N)$, respectively.  
The dynamics of the network is described
by  \cite{Vas}, :

\bee
{{\partial k(s,N)}\over{\partial N}}&=&\nu_{d\rightarrow e} \hat k(s,N) + \nu_{p\rightarrow e} k^*(s,N) - (\nu_{e\rightarrow d} + \nu_{e\rightarrow p} ) k(s,N)
\label{dif1} \\
{{\partial k^*(s,N)}\over{\partial N}}&=&\nu_{d\rightarrow p} \hat k(s,N) + \nu_{e\rightarrow p} k(s,N) - (\nu_{p\rightarrow d} + \nu_{p\rightarrow e} ) k^*(s,N) \label{dif1_1} \\
\hat k(s,N)&=&N-k(s,N)-k^*(s,N),
\label{dif1_2}
\eee

\ni where  $\nu_{d\rightarrow e}$,  $\nu_{p\rightarrow e}$ depend, as will be shown later, only on the network size
and   $\nu_{d\rightarrow p}(s)$,  $\nu_{p\rightarrow d}(s)$ are given as
\[
\nu_{d\rightarrow p}(s)=\nu_{d\rightarrow e} k(s), 
\ \ \ \nu_{p\rightarrow d}(s)=\nu_{e\rightarrow d}k(s).
\]
The last two equations hold since  if, for example,  a node $i'$ is connected to another node $i$ 
by an edge, potential edges from $i'$ are created to all  of the neighbors of the node $i$.

To simplify eqs. (\ref{dif1}--\ref{dif1_2}) several assumptions reflecting to properties of real social networks were
introduced \cite{Vas}:

\begin{enumerate}
\item  All processes leading to edge deletion are neglected:  $\nu_{e\rightarrow p}=\nu_{e\rightarrow d}=0$.
\item  The transition from a potential edge to an edge has a higher probability of occurrence then the transition from being
disconnected to an edge 
$$
\nu_{p\rightarrow e}={\mu_1 \over N},\qquad \nu_{d\rightarrow e} ={\mu_0\over N^2}, 
$$ 
where $\mu_1>0$ and $\mu_0>0$ are constants (for details see \cite{Vas}).
\item  Terms of order $1/N^2$ are omitted.
\end{enumerate}

\ni Under these conditions the system  (\ref{dif1}--\ref{dif1_2}) turns into 

\bee
{{\partial k(s,N)}\over{\partial N}}&=&{\mu_0\over N}  + {\mu_1\over N} k^*(s,N) \label{dif12}\\
{{\partial k^*(s,N)}\over{\partial N}}&=&{{\mu_0  k(s,N)}\over N}-{\mu_1\over N} k^*(s,N)\label{dif12_1}
\eee

\ni with solution ($k_0$ and $k^*_0$ are positive constants) \cite{Vas}

\bee
 k(s,N)=k_0 \left ({N\over  s}\right )^\beta 
\label{dif12sol} \\
k^*(s,N)=k_0^*\left ({N\over s}\right )^\beta, 
\label{dif12sol_1}
\eee
where
\be\label{beta1a}
\beta={\mu_1\over 2}\left (-1+\sqrt {(1+4{\mu_0\over \mu_1}}\right ).
\ee

\ni The  degree distribution of the system is therefore power law (\ref{degdist1}) with the exponent $\gamma = 1+{1\over \beta}$  \cite{AB, DM3}.

\section{ Analysis of the CNN  model}

In what follows  we suggest a more accurate analysis  of the original system (\ref{dif1}--\ref{dif1_2}) describing the
dynamics of the CNN network. From the  three simplifying assumptions described above and employed in \cite{Vas} we use only  the first two.
We thus start with the system of differential equations:

\bee
{\partial k(s,N) \over \partial N} &=& {\mu_0\over N^2} \left (N-k(s,N)-k^*(s,N)\right ) + {\mu_1\over N }k^*(s,N)
\label{dif2}\\
{\partial k^*(s,N)\over \partial N} &= &{\mu_0\over N^2} k(s,N) \left(N-k(s,N)-k^*(s,N)\right) -  {\mu_1\over N}k^*(s,N).\label{dif2_1}
\eee

\ni Unlike in \cite{Vas}, our aim here is a qualitative analysis of this nonlinear system. We would like to
describe asymptotic behavior of solutions of (\ref{dif2}-\ref{dif2_1}) in the space of parameters
$\mu_0$, $\mu_1$. The goal of the following series of substitutions is to transform the system
(\ref{dif2}-\ref{dif2_1}) to a more convenient form, which can be approximated by an autonomous  nonlinear system.

In the continuum approach \cite{{AB, DM3}}   $N$ is regarded a continuous 
variable and $s$ is a node label. First set $N=e^{\tau}$ and substitute 

\bee
x(s,\tau) &=& k(s,\tau) +k^*(s,\tau) \label{subs1} \\
y(s,\tau)&=& k(s,\tau)+1, \label{subs1_1}
\eee

\ni  to the system (\ref{dif2}--\ref{dif2_1}). We obtain 

\bee
x'&=&\mu_0 (1-x e^{-\tau}) y
\label{dif3}\\
y'&=&\mu_0 (1-x e^{-\tau})+\mu_1(x-y+1),\label{dif3_1} \eee

\ni where 
$x(s,\tau)$ and $y(s,\tau)$ are abreviated as $x$ and $y$, respectively and we write $x'$ and $y'$ for ${\partial x(s,\tau) \over \partial \tau}$
and
${\partial y(s,\tau) \over \partial \tau}$, respectively.

Using another substitution

\bee
u &=& 1-x e^{-\tau} \label{subs2} \\
v &=& y e^{-\tau},\label{subs2_1} \eee

\ni together with abbreviations $u(s,\tau)=u$  and $v(s,\tau) = v$, we again get a
new system of differential equations:

\bee
u' &=& 1- u -\mu_0 u v \label{dif4} \\
v' &=& \mu_1 - \mu_1 u - \mu_1 v - v +\mu_0 u e^{-\tau} + \mu_1
e^{-\tau},\label{dif4_1} \eee

\ni with obvious interpretation of $u'$ and $v'$.
It is natural to assume that the function $u(s,\tau)$ (\ref{subs2}) is  bounded\footnote{As the network size $N= e^\tau$ grows, the degrees (both actual and potential) of any node $s\le N$ cannot grow faster than $c \cdot N$ for some positive constant $c$.}. 
Then for $\tau \rightarrow\infty$, the functions $\mu_0 u e^{-\tau}$,
$\mu_1 e^{-\tau}$ tend to zero.  We approximate the  system (\ref{dif4}--\ref{dif4_1}) by an autonomous one

\bee
u' &=& 1- u -\mu_0 u v \label{dif5}\\
v' &=& \mu_1 - \mu_1 u - \mu_1 v- v. \label{dif5_1} \eee

\ni The correctness  of the approximation (\ref{dif5}-\ref{dif5_1}) can also  be  shown if we introduce a new function $z=e^{-\tau}$
and rewrite the system (\ref{dif4}--\ref{dif4_1}) into  a three dimensional one

\bee
u' &=& 1- u -\mu_0 u v \label{dif51}\\
v' &=& \mu_1 - \mu_1 u - \mu_1 v - v+\mu_0 u z + \mu_1 z. \label{dif51_1} \\
z'&=&-z. \label{dif51_2} \eee

\ni Now, we can see from (\ref{dif51}-\ref{dif51_2}) that all of the important  dynamics occurs in the $(u,v)$ plane.

The system (\ref{dif5}--\ref{dif5_1}) possesses two fixed points

\be\label{A}
A=[1,0]
\ee

\ni and

\be\label{B}
B=\left[{{\mu_1+1}\over {\mu_0 \mu_1}},
{{\mu_0\mu_1-\mu_1-1}\over {\mu_0 \mu_1+\mu_0}}\right].
\ee

\ni Having identified the fixed points, the next step is to characterize their stability type. To that end we  linearize  
(\ref{dif5}--\ref{dif5_1}). The linearization matrix $M$ of (\ref{dif5}--\ref{dif5_1}) reads

\be\label{lin1}
M=\left( \begin{array}{cc} -1-\mu_0 v & -\mu_0u \\
          -\mu_1 & -\mu_1-1  \end{array} \right)
\ee

\ni and at the fixed point   $A$  (\ref{A}) we get 

\be\label{lin1A}
M_A=\left( \begin{array}{cc} -1 & -\mu_0 \\
          -\mu_1 & -\mu_1-1  \end{array} \right).
\ee

\ni Analogously at the fixed point $B$ (\ref{B}) we get 

\be\label{lin1B}
M_B=\left( \begin{array}{cc}  - {\mu_0\mu_1\over\mu_1+1}& -{\mu_1+1\over\mu_1} \\
          -\mu_1 & -\mu_1-1  \end{array} \right)
\ee

\ni The types of fixed points $A$, $B$ depend on eigenvalues of 
$M_A$, $M_B$. General form of eigenvalues of the matrix $M$ (\ref{lin1}) is

\be\label{eigA}
 \lambda_{1,2}={Tr (M)\over 2} \pm \sqrt{\left( {Tr(M)\over 2 }\right)^2 - Det (M)},
\ee

\ni where $Tr (M)$ and $Det (M)$ denote trace and determinant of  $M$, respectively. Clearly

$${Tr (M_A)\over2}=-1-{\mu_1\over2}, 
\qquad Det( M_A)=1+\mu_1-\mu_0\mu_1$$
and
$${Tr (M_B)\over2}=-1-\mu_1-{\mu_0\mu_1\over\mu_1+1},
\qquad Det (M_B)=-1-\mu_1+\mu_0\mu_1.$$

Both traces $Tr(M_A)$ and $Tr(M_B)$ are  negative for each $\mu_0>0,\ \mu_1>0$ (which, of course, is fulfilled for the CNN
model), and
the determinants have opposite signs, $Det( M_A)=- Det( M_B)$. 
It therefore follows from (\ref{eigA})
that the sign of the expression
$1+\mu_1-\mu_0\mu_1$ is decisive for the stability type of $A$ and $B$. In the case that  $1+\mu_1-\mu_0\mu_1>0$,
the eigenvalues
of $M_A$ are both negative and $A$ is an asymptotically stable fixed point called sink. The
eigenvalues of $M_B$ have opposite signs and $B$ is an unstable fixed point (saddle).
In the complementary case, $1+\mu_1-\mu_0\mu_1<0$, $A$
and $B$ change their stability types, now $A$ is a saddle and $B$ is a stable fixed point (sink).
The critical case occurs when $1+\mu_1-\mu_0\mu_1=0$. In that case the fixed points $A$, $B$
coincide.

This type  of stability change is called a transcritical
bifurcation (e.g. \cite{GuH, HaK}). In the CNN model the bifurcation occurs when the parameters
$(\mu_0, \mu_1)$
cross the line

\be\label{bifline}
\mu_0=1+{1\over\mu_1}
\ee

\begin{figure}[ht]
\begin{center}
\epsffile{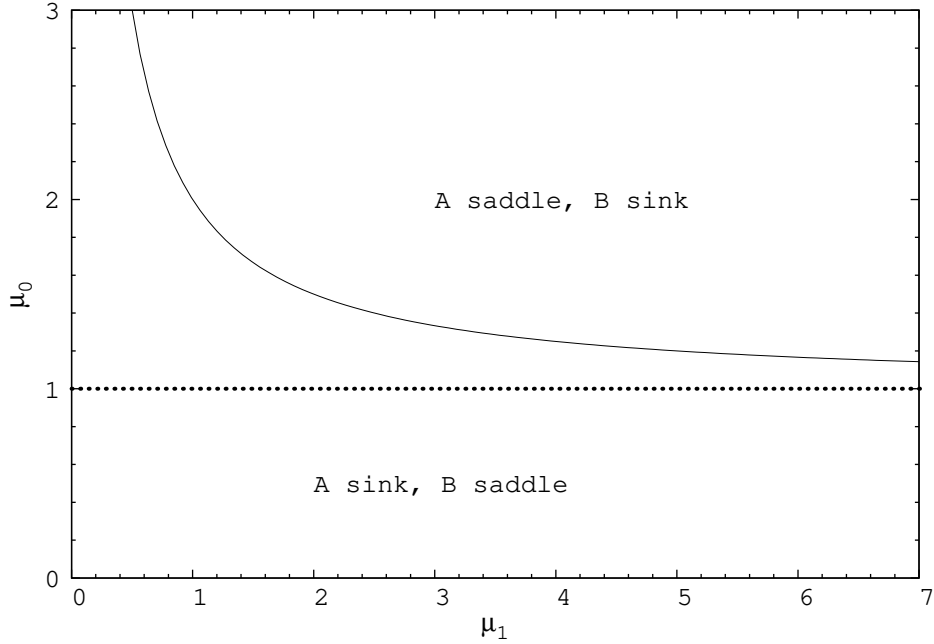}
\caption{Bifurcation line obtained from (\ref{bifline}) in the parameter space of the CNN model delimiting regions of different stability types of fixed points of (\ref{dif2}--\ref{dif2_1}). Dotted horizontal line indicates an asymptotic behavior when $\mu_1\to \infty$. }
\label{Fig.1}
\end{center}
\end{figure}

\ni in the parameter space (see Fig. \ref{Fig.1}).

Let us now consider both cases in greater detail.  In the first case,  $1+\mu_1-\mu_0\mu_1>0$, $A$ is a stable sink and
the eigenvalues of $M_A$ are

\be\label{eigenval1}
\lambda_{1,2}=-1-{\mu_1\over2}\pm\sqrt{\mu_0\mu_1+{\mu_1^2\over 4}}.
\ee

\ni The corresponding eigenvectors of $M_A$ for eigenvalues $\lambda_1$, $\lambda_2$ read

\bee
\vec v_1&=&\left( \begin{array}{c}  -\mu_0 \label{eigenvec1}\\
          -{\mu_1\over 2} +  \sqrt{\mu_0\mu_1+{\mu_1^2\over 4}} \end{array} \right)\\
\vec v_2&=&\left( \begin{array}{c}  -\mu_0 \\
          -{\mu_1\over 2} -  \sqrt{\mu_0\mu_1+{\mu_1^2\over 4}} \end{array} \right).\label{eigenvec1_1}
\eee

\ni The general solution of the linearization which approximates
solutions  of the system (\ref{dif5}--\ref{dif5_1}) and therefore also solutions
of (\ref{dif4}--\ref{dif4_1}) near the stable fixed point $A$ is

\be\label{sol11} \left( \begin{array}{c} u \\ v \end{array} \right)
=\left( \begin{array}{c} 1 \\ 0 \end{array} \right) + c_1 \vec v_1
e^{\lambda_1 \tau} + c_2 \vec v_2 e^{\lambda_2 \tau}. \ee

\ni where $c_1$, $c_2$ are constants. Returning to the original
variables, neglecting $e^{\lambda_2 \tau}$ terms   and using initial
condition $k(s,s)=1$ we get the final approximation

\bee
k(s,N)&=&2 \left({N\over s}\right )^{\beta}-1 \label{solfin} \\
k^*(s,N)&=&2\left( {\mu_0\over \beta} -1 \right) \left({N\over s}\right )^{\beta} + 1, \label{solfin_1}
\eee

\ni where the exponent $\beta$ is given by

\be\label{beta}
\beta = {\mu_1 \over 2} \left(-1+\sqrt{1+{4\mu_0\over { \mu_1}}} \right ).
\ee

\ni Equation (\ref{solfin}) allows us to calculate the
average degree as

\be \label{averA} 
\bar k(N)={1\over N}\sum_{s=1}^{N}k(s,N)={1\over
N}\left[\sum_{s=1}^{N}2 \left({N\over s}\right)^{\beta}-1\right].
 \ee

\ni Approximating the sum in (\ref{averA}) by definite integral,
the asymptotic (as $N\to \infty$) average degree reads

\be \label{averAas}
\bar k={1+\beta\over 1-\beta}. \ee

\ni There is another way of obtaining the asymptotic average
degree $\bar k$. From (\ref{solfin_1}),
\be\label{kstar}
 k^*(s,s)=2 {\mu_0\over \beta} -1 =\sqrt{1+{4\mu_0\over { \mu_1}}}.
 \ee

\ni This result has an interesting interpretation. If a node comes and links itself  to a node $s'$ by one edge, potential edges are
immediately created to all of the neighbors of $s'$. Therefore $k^*(s,s)$ can be interpreted as an asymptotic average degree. The
difference between (\ref{averAas}) and (\ref{kstar}) is due to the neglected terms in (\ref{solfin}), but both calculations lead to the
finite average degree $\bar k$.

In the other case, when $1+\mu_1-\mu_0\mu_1<0$, fixed point $B$ is asymptotically stable
and the eigenvalues of $M_B$ are

\be\label{eigenval2} \lambda_{3,4}=-{1+\mu_1\over
2}-{{\mu_0\mu_1}\over{2(\mu_1+1)}}\pm\sqrt{\left({{1+\mu_1}\over2}+{{\mu_0\mu_1}\over{2(\mu_1+1)}}\right)^2-\mu_0\mu_1+\mu_1+1}
\ee

\ni Clearly $\lambda_4<\lambda_3<0$.

\ni The general solution  approximating solutions  of the system (\ref{dif4}--\ref{dif4_1}) near the
stable fixed point $B$ is therefore

\be\label{sol1} \left( \begin{array}{c} u \\ v \end{array} \right)
=\left( \begin{array}{c} {\mu_1+1\over\mu_0\mu_1}
\\{\mu_0\mu_1-\mu_1-1\over\mu_0\mu_1+\mu_0} \end{array} \right) + c_3 \vec
v_3 e^{\lambda_3 \tau} + c_4 \vec v_4 e^{\lambda_4 \tau}, \ee

\ni where $c_3$, $c_4$ are constants and $\vec
v_i=(v_{i1},v_{i2})^T$,  $i \in \{3,4\}$ are eigenvectors of
$M_B$. Let us denote $\mu={\mu_0\mu_1-\mu_1-1\over\mu_0\mu_1+\mu_0}. $

Backward substitutions lead to the solution in the original variables
$k,k^*$:

\bee
k(s,N)&=&\mu N-c_3 v_{32} N^{1+\lambda_3}- c_4 v_{42} N^{1+\lambda_4}-1
\label{solfin1}\\
k^*(s,N)&=&{\mu\over\mu_1}N-c_3 ( v_{31}- v_{32})
N^{1+\lambda_3}- c_4( v_{41}- v_{42}) N^{1+\lambda_4}+1
\label{solfin1_1} \eee

\ni where coefficients $c_3$ $c_4$ are $s$ -dependent.

Since $\lambda_4<\lambda_3<0$, the largest exponent of $N$ is $1$ and we can approximate the solution by

\bee
k(N)&=&\mu N \label{approx1} \\
k^*(N)&=&{\mu\over\mu_1}N \label{approx1_1} \eee

\ni For  $N\to\infty$,

\be \lim_{N\to\infty}{k(s,N)\over k^*(s,N)}=\mu_1,
\ee

\ni and $\mu_1$ is thus an asymptotic ratio of the node's degree and its
potential degree. 

It is possible to find a neater approximation. As
$\lambda _4<-1$, neglecting only terms with $N^{1+\lambda_4}$ in
(\ref{solfin1}) we obtain

\bee
k(s,N)&=&\mu N-c_3 v_{32} N^{1+\lambda_3}-1\label{approx2}\\
k^*(s,N)&=&{\mu\over\mu_1}N-c_3 ( v_{31}- v_{32})
N^{1+\lambda_3}+1 \label{approx2_1} \eee

\ni which together with the initial condition $k(s,s)=1$ leads to the final result

\bee k(s,N)&=&\mu\left (1-\left(N\over
s\right)^{\lambda_3}\right)N+
2 \left(N\over s\right)^{1+\lambda_3}-1
\label{approx3} \\
k^*(s,N)&=&{\mu\over\mu_1}\left(1+\left(N\over
s\right)^{\lambda_3}\right)N-2 \left(N\over
s\right)^{1+\lambda_3}-c_3 v_{31} N^{1+\lambda_3}+1 
\label{approx3_1}
\eee where

\be\label{vec31} 
v_{31}=1+{1\over \mu_1}. 
\ee

\ni Again, from (\ref{approx3}) we can derive the
average degree:

\be \label{averB} 
\bar k(N)={1\over N}\sum_{s=1}^{N}k(s,N)={1\over
N}\sum_{s=1}^{N}\mu\left (1-\left(N\over
s\right)^{\lambda_3}\right)N+ 2 \left(N\over
s\right)^{1+\lambda_3}-1 .\ee

\ni Approximation of the sum in (\ref{averB}) by definite integral and  taking the limit
$N\to \infty$ leads to

\be \label{averBas}
\bar k=\infty. 
\ee

\ni As before, the same asymptotic result can be obtained from

 \be\label{approx4}
k^*(s,s)={\mu\over\mu_1}2s-1 -c_3 v_{31} s^{1+\lambda_3} \ee

\ni and

\be \label{approx5}
\lim_{s\to{\infty}}k^*(s,s)=\bar k^*=\infty.\ee

\ni The
asymptotic average degree tends to infinity when the
fixed point $B$ is stable.

To conclude our analysis we can state  that the CNN model
(\ref{dif1}-\ref{dif1_2}) exhibits two different types of behavior, deliminated by the
bifurcation line (\ref{bifline}) in the parameter space $(\mu_0,\mu_1)$.

For $\mu_0<1+{1\over\mu_1}$ the degree  $k(s,N)$ with increasing $N$ grows sublinearly with 
  the exponent $\beta$ given by (\ref{beta}), where $0<\beta<1$.
Due to  approximations (\ref{averAas}), (\ref{kstar}) the asymptotic
average degree is in this case constant. The degree distribution is
a power law with exponent $\gamma=1+{1\over\beta}$ \cite{AB,DM3}.

For $\mu_0>1+{1\over\mu_1}$, the growth of the degree $k(s,N)$ is
linear (exponent $\beta=1$) with increasing $N$. Approximations (\ref{averBas}),
(\ref{approx5}) imply that now the asymptotic average degree tends
to infinity. The degree distribution is a power law with the exponent
$\gamma=2$.

\begin{table}[ht]
\caption { $\gamma$ exponent of the degree distribution in a parameter space region where the fixed point A is a sink and B is a saddle. $\mu_0=1$, $\gamma_t$ is calculated with a help of (\ref{beta1a},   \ref{beta}) and $\gamma_n$ is a numerical value.}
\vskip 0.5truecm
\centering
\begin{tabular} {c c c}
\hline\hline
$\mu_1$ & $\gamma_t$ & $\gamma_n$\\
\hline
0.25 & 3.56 & 3.49 \\
0.5   & 3.0   & 3.03 \\
1.0   & 2.62 & 2.66 \\
\hline
\end{tabular}
\label{Table 1.}
\end{table}

We supported our analytical results by the numerical simulations. We simulated the model for parameters $\mu_0$, $\mu_1$ in both regions of the parameter space  (Fig. \ref{Fig.1}), using the equations  (\ref{dif2}, \ref{dif2_1}). Our network has 100000 nodes. In the first region, where the V\'azquez solution holds, we simulated the model for $\mu_0=1$ and for the three different values of $\mu_1$. The numerical results are in a good accordance with the theoreticaly predicted values (Tab.\ref{Table 1.}). 

\begin{figure}[ht]
\begin{center}
\epsffile{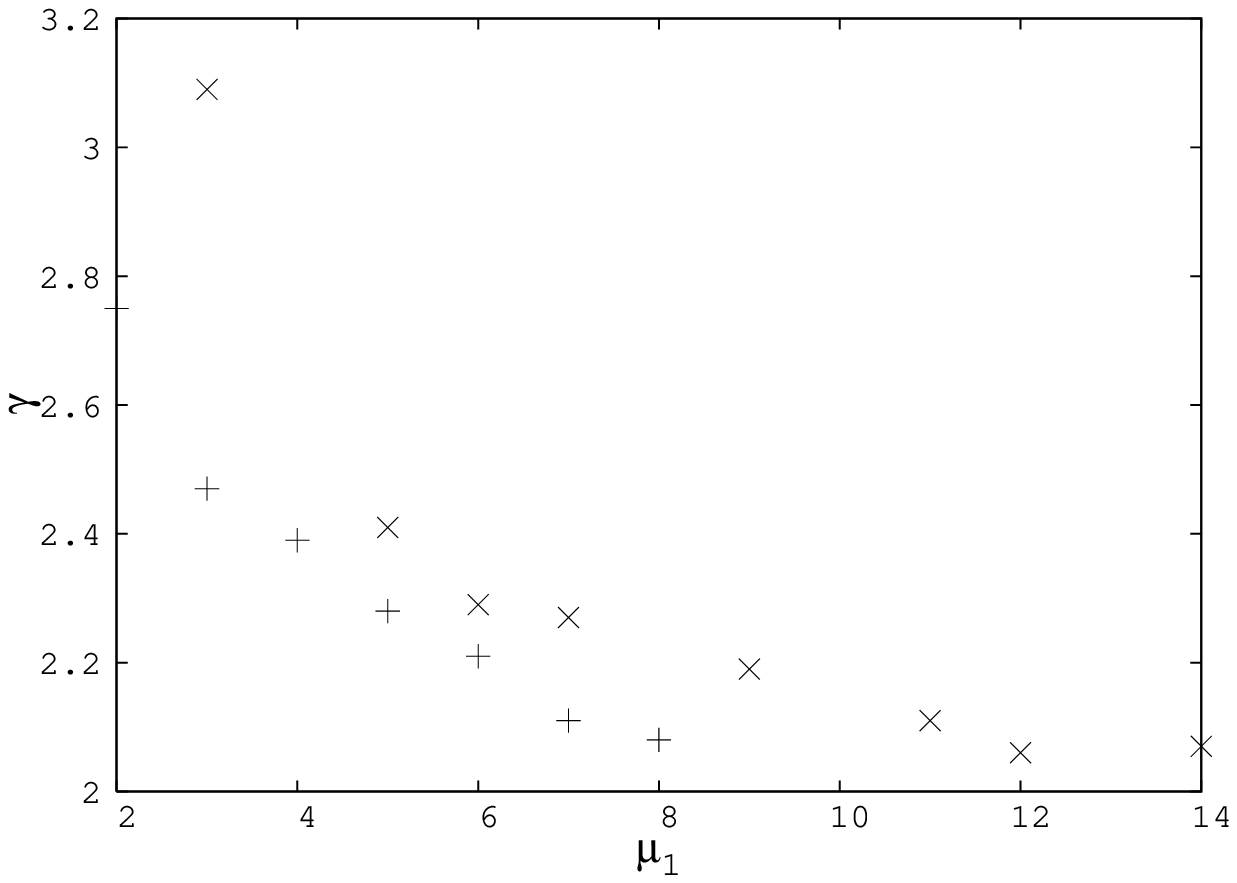}
\caption{Convergence of the $\gamma$ exponent to the predicted value $\gamma=2.0$ in the region above the line of the transcritical bifurcation. $\mu_0=3$ for the upper curve (x) and $\mu_0=2$ for the lower curve (+). All simulations were done for 100000 nodes. }
\label{Fig.2}
\end{center}
\end{figure}

Above the bifurcation line the numerical simulations does not converge very well for the parameter values  $\mu_0$, $\mu_1$ close to the bifurcation line.  We belive, this is partly due to the fact, that the second terms of the equations (\ref{solfin1}) and (\ref{solfin1_1}) influence the convergence. The results for several parameter values are in the (Tab.\ref{Table 2.}). (Fig. \ref{Fig.2}) shows the convergence of the $\gamma$ exponent for $\mu_0=2,3$ and growing $\mu_1$.

\begin{table}[ht]
\caption { $\gamma$ exponent of the degree distribution in a parameter space region where the fixed point A is a saddle and B is a sink.  $\gamma_t$ is a teoretical value of the scaling exponent, and $\gamma_n$ is a numerical value.} 
\vskip 0.5truecm
\centering
\begin{tabular} {c c c c}
\hline\hline
$\mu_0$ & $\mu_1$ & $\gamma_t$ & $\gamma_n$\\
\hline
2 & 6 & 2.0 & 2.21 \\
2 & 8 & 2.0 & 2.08 \\
3 & 9 & 2.0 & 2.19 \\
3 & 12 & 2.0 & 2.06 \\
\hline
\end{tabular}
\label{Table 2.}
\end{table}

\section{Correspondence of the CNN model to other models}

Having obtained analytical solutions for the CNN model, we would like to establish whether some known models 
are, at least for some parameter settings, identical to the CNN model. We have tested a growing variant of the CNN 
model (the GCNN model)  and the recursive search (RS) model, which is derived from the PRW model. All of them are described in \cite{Vas}.

\subsection{The GCNN model}

In \cite{Vas} V\'azquez proposed a growing variant of the CNN model, the GCNN model. Fix some $u \in (0,1)$. The GCNN network evolves by following iteratively performed rules:

\begin{enumerate}
\item With  probability $1-u$  \ a new vertex is introduced and a new edge from the new vertex to a randomly 
selected network node is created. 

\item With the probability $u$ one potential edge (selected at random) is converted to an edge. 

\end{enumerate}

\ni The author expects  that in the regime of large $N$, the CNN and GCNN models  will behave similarly.
It has been stated  
that the evolution rules are consistent with setting $\mu_0=1$ and $\mu_1={u\over{1-u}}$ in 
(\ref{dif12}--\ref{dif12_1}):

\bee
{{\partial k(s,N)}\over{\partial N}}&=&{1\over N}  +\left({u\over{1-u}}\right) {1\over N} k^*(s,N) \label{dif12a} \\
{{\partial k^*(s,N)}\over{\partial N}}&=&{{ k(s,N)}\over N}-\left({u\over{1-u}}\right){1\over N} k^*(s,N),\label{dif12a_1}
\eee

\ni How to interpret this? 
The number of nodes $N$ is now proportional to the time of the network development. If the time is measured by 
coming nodes, it is necessary to rescale $k^*$ by the  factor $u\over {1-u}$ which is, in fact, the ratio of 
transformed potential edges to one node. For our next analysis it is quite correct to deal with the original set 
of equations (\ref{dif12}--\ref{dif12_1}, \ref{dif12a}--\ref{dif12a_1}), for reasons which are to be made clear later.

Having the  parameter settings $\mu_0=1$ and $\mu_1={u\over{1-u}}$   and taking into account that 

\be\label{gamma}
\gamma=1+{1\over\beta},
\ee

\ni  the scaling exponents  $\beta$  and $\gamma$ of the GCNN model are

\be\label{bGCNN}
\beta={u\over{2(1-u)}}\left(-1+\sqrt{1+4{{1-u}\over u}}\right)
\ee

\ni and

\be\label{gamma1}
\gamma(u)=1+{{2(1-u)}\over u} \left( -1 +\sqrt{1+4{{1-u}\over u}}\right)^{-1}.
\ee

\ni The limiting cases, analyzed in \cite{Vas} are 

\be\label{glim1}
\gamma(0)=\infty 
\ee

\ni and

\be\label{glim2}
\gamma(1)=2. 
\ee
 
\ni Viewing these results in the light of our solution, it is clear that the condition $\mu_0=1$ ensures, we are 
in the region of the parameter space, where the fixed point $A$ is an asymptotically stable  sink (Fig. \ref{Fig.1}).
Exponents $\beta$ and $\gamma$ are therefore given by (\ref {bGCNN}) and (\ref{gamma1}).
This is also why it is correct to use the original equation set (\ref{dif12}, \ref{dif12a}), which 
gives a good result in this 
part of the parameter space.  From   (\ref{gamma1}) we have that $\gamma(0)=\infty$.

There is an alternative argument:  If $u=0$, only the node addition rule works,
each new node creating 
a new edge to a randomly 
selected node in the network.
Such network, 
growing by the random node addition,
has been analyzed in \cite{DM1, AB}. In such networks the degree distribution decreases exponentially, which corresponds 
to  $\gamma(0)=\infty$.

To analyze the second limiting case, we rewrite (\ref{bifline}), 

\be\label{bifline1}
\mu_1={1\over{\mu_0-1}}.
\ee

\ni Comparing (\ref{bifline1}) with $\mu_1={u\over {1-u}}$, it is clear, that for ${1\over \mu_1}=0$ and $u=1$ both expressions are identical. 
That means, that for the limiting case $u=1$, we are assymptotcally on the bifurcation line (\ref{bifline}) in the parameter space, where the solution changes to that described by the equations (\ref{approx1}). 
Here the scaling exponents are $\beta=1$ and $\gamma=2$.

Different scaling exponents in the two cases  $\gamma(0)$ and $\gamma(1)$ (\ref{gamma1}) are therefore a natural consequence of the fact, that the solution lies in different regions of the parameter space divided by the line of the transcritical bifurcation (\ref{bifline}).

\subsection{The GCNN model: $u={1\over 3}$ case}

We get another interesting result when ${\beta}={1\over 2}$ (\ref{bGCNN}). Using (\ref{gamma}) leads to $\gamma =3$. This  value  of the $\gamma$ exponent is typical for several well known
complex network models, for example the  Barab\'asi Albert  model (BA model) \cite{BA} or the scale free hierarchical model (SFH model) \cite{My}.
In \cite{My} we established a simple process of network growth that leads to  scale free hierarchical network structure in which  both degree and clustering coefficient distributions are power law  (\ref{degdist1}, \ref{cldist1}). 

\begin{figure}[ht]
\begin{center}
\epsffile{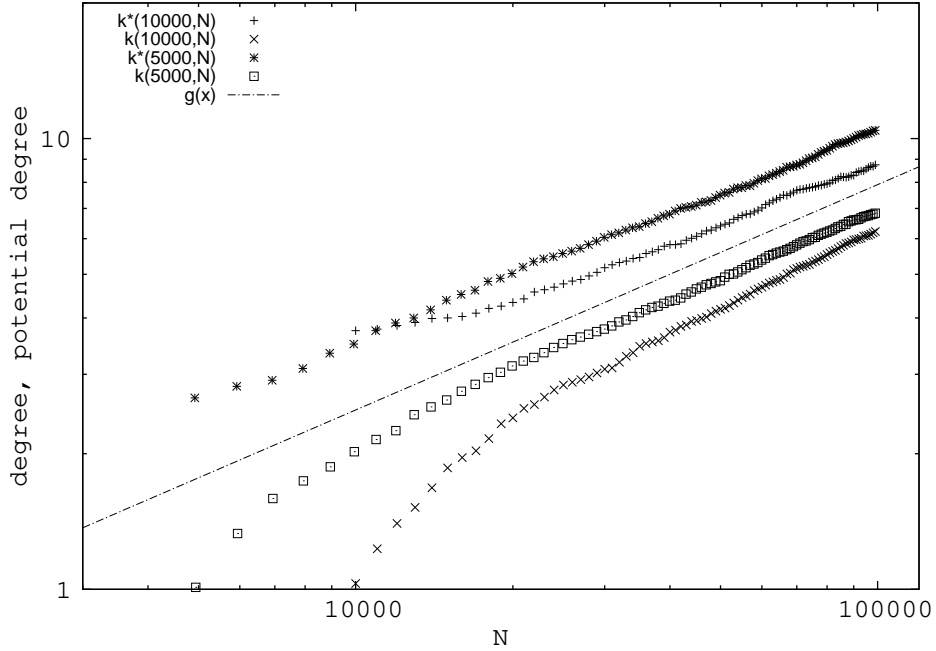}
\caption{Log-log plot of the degree and potential degree developement with the size of the GCNN network for two different nodes ($s=5000,10000$).   $k(s,N)$ ($k^*(s,N)$) is a degree (potential degree) of the node $s$ and network size $N$. The  line  has a slope $1\over 2$ (\ref{knehsol}).}
\label{Fig.3}
\end{center}
\end{figure}

In the GCNN model either new vertex with new edge is added to the system with probability $1-u$, or, with probability $u$, some potential edge is chosen and turned into an edge connecting two nodes with a common third node to which the two nodes are connected.
A triangle is thus created. Node addition causes creation of several new potential edges in the system. 
Let us assume that 

\be\label{cc}
{\partial k^*(s,N)\over \partial N}={\partial k(s,N)\over \partial N},
\ee

\ni holds for the GCNN model for some value of  $u$. Using this assumption, we get from  (\ref{dif12a})

\be\label{khviezd}
{\partial k^*(s,N)\over \partial N}={1\over N} +\left({u}\over{1-u}\right){1\over N} k^*(s,N),
\ee

\ni with the solution 

\be\label{khsol}
k^*(s,N)= \left({N\over s}\right)^{{u}\over{1-u}}c-{{1-u}\over{u}}.
\ee

\ni where the  exponent $\beta\left(u\right)={u\over{1-u}}$ and $c$ is a  constant.  Comparing this value of $\beta$ with (\ref{bGCNN})
 we get that $u={1\over 3}$.  That means, that $\beta\left({1\over 3}\right) ={1\over 2}$ is a result of the fact, that, as the network grows, both
degree and potential degree of nodes change the same way. 

Summing  (\ref{dif12a}) and (\ref{dif12a_1})  and using the assumption  (\ref{cc}) we get 

\be\label{k}
2{\partial k(s,N)\over \partial N}={1\over N}(1+k(s,N))
\ee

\ni with  solution (under initial condition $k(s,s)=m$) (Fig.\ref{Fig.3})

\bee\label{knehsol}
k(s,N)=(m+1) \left({N\over s}\right)^{1\over 2}-1\\
k^*(s,N)=(m+1) \left({N\over s}\right)^{1\over 2}-2. \nonumber
\eee

To conclude this analysis, we can state that if the special  condition $ {\partial k^*(s,N)\over \partial N}={\partial k(s,N)\over \partial N}$ holds in the GCNN model, then $u={1\over 3}$ and the scalings exponents $\beta={1\over 2}$ and 
$\gamma =3.0$. It would be interesting to investigate, whether the GCNN model can be transformed to some other known models having degree distribution governed by the exponent $\gamma =3.0$. We belive, that the scale free hierarchical (SFH) model \cite{My} and the GCNN 
model are similar, but we leave this to future work.

\subsection{Correspondence of the GCNN model to  the RS model}

Recursive search model (RS model) has been investigated by V\'azquez in \cite{Vas}. It is a variant of his  particular 
random walk model (PRW). General random walk model simulates how the structure of a network is discovered by surfers
\cite{Vas}. 
 The model has been inspired  by the manner people obtain information by surfing on the world wide web. 
Here the web pages are nodes, and hyperlinks are edges.

 How does the surfer move? There are two possibilities.  Either to  jump on a page selected randomly, and then follow 
a hyperlink, or to jump on another randomly selected page.  PRW model is a simpler version of this random walk model.
The dynamics of the PRW model consists of adding and walking.  The process starts with a single node. Adding means, that the
 new vertex is created and connected to the one of the existing vertices by an edge. Walking is defined as follows: if an edge is linked to a vertex in the network, the edge is also created to one of its nearest neighbors with probability  $q_e$. When no edge is created, adding rule is applied. 

 In the RS model  the walking rule is slightly changed \cite{Vas}. Here walking means that if an edge is created to a vertex in the network, then with probability $q_e$ an edge is also created to all of its nearest neighbors. When no edge is created the adding rule is applied  with the  probability $1-q_e$. 

V\'azquez \cite{Vas} solved the RS model analytically for $q_e=0$ and $q_e=1$. In the first case ($q_e=0$) only 
adding is present. No link is given  to the neighbors of the chosen vertex.  We believe that this is the same situation as 
in the GCNN model with $u=0$.  Here adding creates new potential edges, but none of them is turned to  an edge. 
It has been shown, that the degree distribution in the recursive search model decreases exponentially, 
which corresponds to the $\gamma(0)=\infty$ (\ref{gamma1}) case in the GCNN model. And, again, we can say, that the network dynamics is the same  as  that of the network 
growing by the random node attachment \cite{DM1, BA}.

In the second limiting case  $q_e$ is close to $1$ and  adding in the RS model is almost not present. But if a node is eventually added, link is created to a randomly chosen node in the network and ``potential edges'' among the new node and all of the neighbors of
chosen node are created and immediately converted  to edges. This corresponds to the GCNN model with $u=1$. Each time step a potential edge is transformed to an edge, with almost no adding, all potential edges are finally transformed into edges. Degree distribution for both models is a power law with scaling exponent $\gamma=2$.

\section{Conclusion}

We present a detailed analysis of the CNN network model \cite{Vas}. The 
parameter space 
is  divided  into two regions in which the CNN model has different behaviour (Fig.\ref{Fig.1}). The degree distribution in both regions is a power law, 
but the scaling exponent $\gamma$ is different in different parts of the parameter space. We supported our analytical studies by the numerical simulations. While the results belove the bifurcatin line (Fig.\ref{Fig.1}) closely follow the predicted values  (Table.1), above the line the convergence to the predicted value $\gamma=2.0$ is worse in the proximity of the bifutcation line  (Table.2), (Fig.\ref{Fig.2}). We have nevertheless shown, that the $\gamma$ exponent converges to the theoretically predicted value.

Our theoretical analysis enabled us to understand the dynamics of the growing variant of the CNN model, the GCNN model, in two limiting cases of $u=0$ and 
$u=1$. The two cases belong to the different regions of the CNN parameter space.

We analysed the GCNN model with $u={1\over 3} $, in which case the scaling exponent $\gamma$ of the degree distribution is $\gamma=3.0$. We have found, that this results from the fact, that both the degree and the potential degree of nodes in the 
GCCN network changes similarly. We speculate that the same holds for the SFH model and that the two models are related when 
$u={1\over 3}$. 

Moreover,  we have shown that the RS model and GCNN model in the limiting cases $u=0,1$ and $q_e=0,1$ have similar dynamics.

\section*{Acknowledgemets}
Decisive part of this work has been done under the support of the Ramsay - Yunaby research grant at the 
School of Computer Science, University of Birmingham.
This work has been also supported by VEGA grants $1/0476/11$ and $2/0019/10$. All numerical network simulations were analysed with a help of the Network Workbench tool \cite{NWB}.

We are also gratefull to our collegue Dr Peter N\'ather for his advices.

\end{document}